\documentclass[aps,prb,reprint,groupedaddress]{revtex4-2}

\usepackage[english]{babel}

\usepackage{amsmath}
\usepackage{graphicx}
\usepackage[colorlinks=true, allcolors=blue]{hyperref}
\usepackage{float}
\usepackage{xcolor}
\usepackage{color, colortbl} 
\usepackage{diagbox}
\usepackage{soul}

\definecolor{Gray}{gray}{0.9}

\input{alphabet}

\newcommand*{\colorboxed}{}
\def\colorboxed#1#{%
  \colorboxedAux{#1}%
}
\newcommand*{\colorboxedAux}[3]{%
  \begingroup
    \colorlet{cb@saved}{.}%
    \color#1{#2}%
    \boxed{%
      \color{cb@saved}%
      #3%
    }%
  \endgroup
}

\begin{document}

\title{Inhomogeneous  probes for BCDI: Toward the imaging of dynamic and distorted crystals}

\author{I. Calvo-Almazán}
\affiliation{Aragon Nanoscience and Materials Institute, CSIC - University of Zaragoza, Calle de Pedro Cerbuna 9, 50009 Zaragoza, Spain.}

\author{V. Chamard}
\author{T. Gr\"unewald}
\author{M. Allain }
\affiliation{Aix-Marseille Universit\'e, CNRS, Centrale Marseille, Institut Fresnel UMR7249, 13013 Marseille, France}

\date{\today}


\begin{abstract}
    This work proposes an innovative approach to improve Bragg coherent diffraction imaging (BCDI) microscopy applied to time evolving crystals and/or non-homogeneous crystalline strain fields, identified as two major limitations of BCDI microscopy. Speckle BCDI (spBCDI), introduced here, rests on the ability of a strongly non-uniform illumination to induce a convolution of the three-dimensional (3D) frequency content associated with the finite-size crystal and a kernel acting perpendicularly to the illumination beam. In the framework of Bragg diffraction geometry, this convolution is beneficial as it encodes some 3D information about the sample in a single two-dimensional (2D) measurement, \emph{i.e.}, in the detector plane. With this approach, we demonstrate that we can drastically reduce the sampling frequency along the rocking curve direction and still obtain data sets with enough information to be inverted by a traditional phase retrieval algorithm. 
    Numerical simulations, performed for a highly distorted crystal, show that spBCDI allows a gain in the sampling  ratio ranging between 4 and 20 along the rocking curve scan, for a speckle illumination with individual speckle size of 50 nm. Furthermore, spBCDI allows working at low intensity levels, leading to an additional gain for the total scanning time. Reduction of a factor of about 32 were numerically observed. Thus, measurements in the 0.3 s time scale at 4th generation synchrotrons become feasible, with a remarkable performance for the imaging of strongly distorted crystals. Practical details on the implementation of the method are also discussed. 
    \end{abstract}

    \maketitle

\section*{Introduction}
\label{sec:intro}
\indent%
BCDI is a 3D microscopy approach based on the numerical processing of a series of diffracted intensities, produced by a 3D finite-size crystalline sample illuminated by a coherent x-ray beam and sampled, at least, at twice the Nyquist frequency, referred to as over-sampling \cite{Robinson:2009aa,Pfeifer:2006aa,Cha:2016aa,Miao:2000aa}. 
The numerical processing consists in an iterative phase retrieval from the the diffracted intensity information, which leads to a 3D complex-valued map of the sample, where the modulus corresponds to the electronic distribution for a specific family of lattice planes while the phase encodes information about any deviation of these lattice planes from a reference, perfectly periodic crystal \cite{Robinson:2009aa}. Hence, BCDI is particularly sensitive to crystalline distortions (\emph{e.g.}, strain fields, dislocations, lattice rotation, \emph{etc}.) and is capable of providing 3D images of crystalline samples with size ranging from hundreds of nanometers up to a couple of micrometers, with spatial resolution in the order of a few tens of nanometers \cite{Pfeifer:2006aa,Robinson:2009aa}.\\

\indent%
Hard x-rays (with energy larger than 8 keV) eases the implementation of BCDI for \emph{in situ} experiments, with exciting applications in a wide variety of scientific fields including energy-related materials (e.g., catalysis, batteries \cite{Atlan:2023th,Ulvestad:2015aa,Choi:2020vp}), environment (e.g., mineral dissolution  \cite{Yuan:2019tf,Clark:2015aa}) or biomineralisation \cite{Ihli:2019aa}, only to cite a few. However, BCDI fails at imaging crystals with strongly non-homogeneous crystalline strain fields, limiting its scope to crystals with rather narrow strain distribution \cite{Robinson:2020tw,Huang_Robinson:2011aa}. Additionally, the data acquisition process involves the collection of a series of 2D diffraction patterns with a planar detector, along the rocking curve (RC) direction. It corresponds to the sampling of the 3D Bragg peak intensity distribution, performed with an angular scan of the sample about the axis perpendicular to the diffraction plane \cite{Maddali_Li:to5203} and at a sampling frequency at least twice the Nyquist frequency along this direction \cite{Xiong:2014aa}. This time-consuming data acquisition process affects directly the temporal resolution of 3D BCDI, which is rather poor as the sample must remain static during the time needed to fully explore the 3D intensity distribution of the crystal \cite{Ulvestad:2016aa}. This limitation precludes the imaging of  dynamic systems, especially for highly distorted crystals, which require a very high degree of oversampling, \emph{i.e.}, increasing the number of measurements along the RC scan \cite{Wu:cw5029}. In this work, we propose a new approach, called \emph{speckle} BCDI (spBCDI), to boost the measurement efficiency and enable the imaging of dynamic and highly distorted crystals.  \\

\indent%
Following the progress in optical microscopy \cite{Chen:2023uz,Mudry:2012aa,Gustafsson:2000aa,Fiolka:2019aa} and inspired by several implementations in the x-ray regime \cite{Zhao:2023aa,Zhang:2016us,Maiden:2013ww,Morrison:18,GuizarSicairos:2012aa}, we propose to manipulate the coherent x-ray \emph{incident} wavefront with a \emph{modulator phase plate} (MPP) to perform BCDI with a speckled, strongly non-uniform beam. In reciprocal space, the intensity distribution produced by a 3D crystalline sample under a plane-wave illumination is described as the square modulus of the diffracted field, itself given by the 3D Fourier Transform (FT) of the sample, for measurements performed in the Fraunhofer regime \cite{Robinson:2009aa,Pfeifer:2006aa}. Under non-uniform illumination, the scattering process induces a 3D convolution of the sample FT by a structured kernel acting along the direction perpendicular to the incident beam \cite{Mudry:2012aa,Li:2021ud}. In the framework of the Bragg diffraction geometry, because the direction of the incoming beam does not align with the exit direction, any diffracted intensity is expected to probe the sample FT \textit{concurrently} at various positions along the RC direction. Thereby, the high structural richness of the speckle illumination is used to increase the 3D information content of each detector slice in the RC direction, thus reducing  the sampling frequency, and hence, the measurement time.   

\indent
Herein, numerical simulations show this effect, which improves for larger incident angle and smaller speckle size. It enables, not only to relax the sampling conditions in a RC scan down to factors ranging between 4 - 20, but also to reduce the signal-to-noise ratio that is required for a succesfull reconstruction.
Thereby, it yields potential reductions of the measurement time up to a factor of about 32 with respect to plane wave BCDI methodology, which translates into total acquisition times of the order of  0.3 s for the whole data-set, at 4th generation synchrotron sources \cite{Borland:2014aaa,Hettel:2014aa,Li:2022wf}. Another important advantage of spBCDI, is the observed robustness of the method with respect to heavily distorted crystals, which allows one to image crystalline samples with phase distributions containing abrupt discontinuities between domains and/or large strain fields. Such unprecedented performances, coupled to the enhanced efficiency of  spBCDI measurements, open exciting experimental avenues, to access the wide range of time scales and strain scales associated to different structural drivers for \emph{e. g.}, crystal dissolution or growth in liquid medium \cite{Teng:2000aa,Dove:2007aa,Dove:2005aa}, crystal damages under x-ray illumination \cite{Mastropietro:2013aa}, nanoparticles annealing \cite{Lauraux:2020aa}, morphological and internal structure changes upon catalysis \cite{Carnis:2021ux}, etc. \\

This article is structured in three main sections. The first section introduces the sort of structural information encoded in reciprocal space when the crystal is illuminated by a plane wave or a speckled beam and its consequences for the sampling conditions. It further provides a qualitative guide to estimate the reduction factor for the sampling along the rocking curve, which is found to be function of the speckle size and the incident angle. The second section numerically addresses the performances of spBCDI regarding several experimental parameters, including various degrees of sampling, signal-to-noise ratio, knowledge of the exact crystal shape and the incident angle. Comparison of the new method with respect to standard BCDI is also presented, evidencing dramatic gains in acquisition time and therefore justifying development at 4th generation synchrotron sources,  whose practical implementation is briefly discussed in the third section.
\begin{figure}[t]
  \begin{center}
    \includegraphics[width = 1.0 \columnwidth]{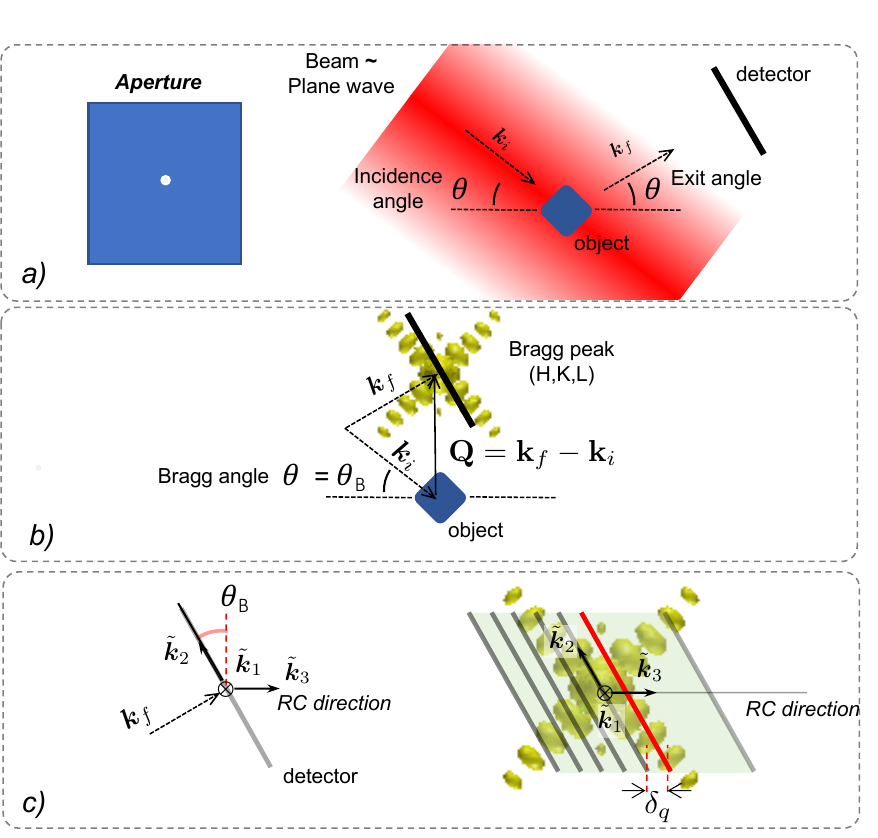}
    \caption{
        Bragg coherent diffraction imaging (BCDI): illumination and geometrical considerations.
        (a) A typical x-ray Bragg coherent diffraction imaging experiment can be
        described as a crystalline sample illuminated by a plane wave
        generated by a point-wise source in the aperture plane, which is located in
        the far-field of the sample.
        (b) The incident  vector $\kb_i$  and the exit vector  $\kb_f$ normal to
        the detector surface are in ``Bragg condition'', when the momentum
        transfer vector $\Qb := \kb_f - \kb_i$ coincides with a point in the reciprocal
        lattice, hence defining a Bragg angle $\theta_{\text{\tiny    B}}$. In that geometry the detector records a 2D slice of the diffracted intensity distributed around the Bragg peak.
        (c) During the RC scan, the rotation of the reciprocal lattice about its origin is equivalent to the sampling of the intensity produced by the sample's 3D Fourier-transform along the RC direction $\tilde{\kb}_3$. Because the RC direction is non-orthogonal to the 
        detector surface, the measurement is naturally described    
        within a non-orthogonal frame $(\tilde{\kb}_1,\tilde{\kb}_2,\tilde{\kb}_3)$. 
        }
    \label{fig:BCDI_geometry}
  \end{center}
\end{figure}
%

\section{Information encoding \textit{via} non-uniform illuminations}
\label{sec:recip_space_info}


\indent%
BCDI is a 3D lens-less microscopy approach aiming to solve the \emph{phase problem}, i.e., retrieve the phase of a complex-valued field, from the measurement of its intensity. It requires the fine sampling (or oversampling) of the 3D diffracted intensity produced by a nano- or micro-sized crystal illuminated with a plane wave \cite{Miao:1998aa,Sayre:1952aa}.
Perpendicularly to the exit beam, the oversampling is ensured by the 2D-detector pixel size (Fig.~\ref{fig:BCDI_geometry}), whose angular resolution can be incremented by increasing the sample-to-detector distance \cite{Robinson:2009aa}, a question which is not further addressed in this work. To probe the third direction, an angular scan is usually performed \cite{Maddali_Li:to5203}. It consists in the acquisition of a series of diffraction patterns measured while the crystal is illuminated at different orientations, such that the detector plane records the intensity distribution around a specific point of the periodically structured reciprocal space, \emph{i.e.}, a Bragg peak (Fig.~\ref{fig:BCDI_geometry}$b$, $c$). The angular scan often corresponds to the rocking curve scan, so that the crystal is tilted about an axis perpendicular to the diffraction plane (Fig.~\ref{fig:BCDI_geometry}c) \cite{Li:to5204,Maddali_Li:to5203}. However, other 3D reciprocal space sampling schemes can be designed, introducing different angular or  energy scans \cite{Cha:2016aa,Maddali_Li:to5203,Li:to5204,Li:2021ud}. Although we did not further explore the other scanning modalities, similar gain should be obtained as spBCDI generates 3D information in the 2D plane of the detector, even at fixed angular position.  \\ 

\begin{figure*}[t]
\begin{center}
\includegraphics[width=1.\textwidth]{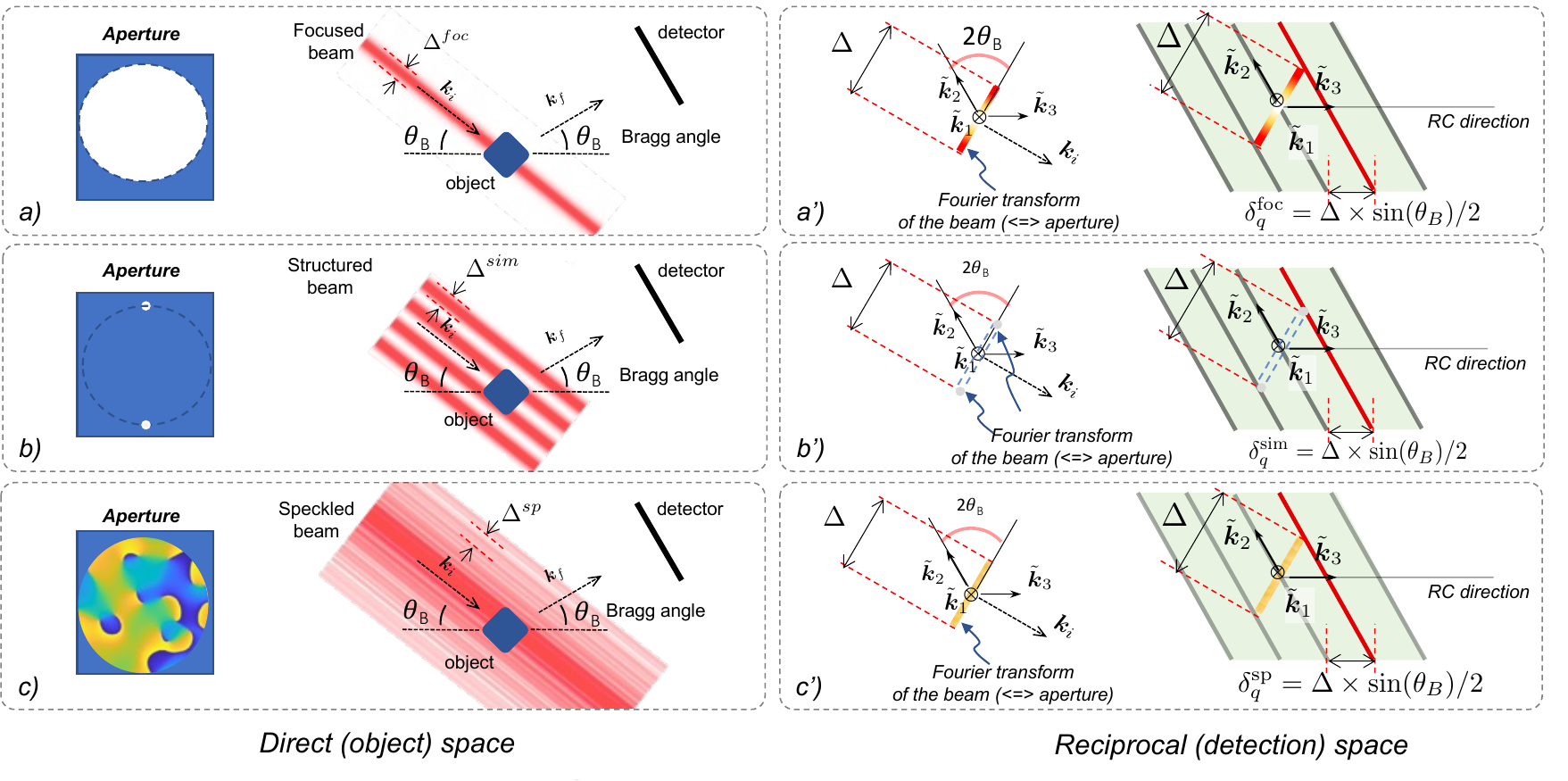}
\caption{%
Schematics of microscopy modalities using inhomogeneous probes: (a) a focused beam, 
(b) an harmonic structured illumination, and (c) a speckled beam. 
Each inhomogeneous probe is characterized by its complex transmission function 
within the optical aperture. The numerical aperture sets an identical bandwidth 
$\Delta$ for each kernel $\mathcal{F}P$ spreading $\mathcal{F}\rho$ in reciprocal 
space. }
\label{fig:general_scheme}
\end{center}
\end{figure*}

The oversampling ratio (OSR) provides an estimate of the sampling step $\delta_q$ required to produce an image of an object with size $\Delta_r$ in the conjugated direction of reciprocal space:
\begin{equation}
\sigma \equiv \frac{1}{\delta_q \Delta_r}. 
\label{eq:OSR_stBCDI_recipspace}
\end{equation} 
\indent%
The Shannon-Nyquist theorem establishes that $\sigma \geq 2$ for all the three dimensions of reciprocal space to produce an invertible data set, \emph{i.e.}, a set of intensities which can be inverted by a phase retrieval algorithm to produce an (essentially unique) image of the complex-valued sample (modulus and phase)  \cite{Sayre:1952aa}.  
Eq. \ref{eq:OSR_stBCDI_recipspace} enables to define an OSR along the RC direction $\widetilde{\kb}_3$, $\sigma_{RC}$, which will depend on the sampling step $\delta_q$ determining the distance between contiguous detector slices in a RC scan. It can be connected to specific experimental parameters since: $\delta_q = |\mathbf{G}_{HKL}|\delta\theta$, where $\delta\theta $ is the angular shift by which the sample is rocked for each acquisition along the RC scan and where $\mathbf{G}_{HKL}$ is the Bravais lattice vector associated to the $HKL$ planes family in the reciprocal space. Thus, complying with the Shannon-Nyquist criterion along the RC imposes that $\sigma_{RC} \geq 2$ which leads to the adjustment of the angular step with respect to the object size. In practice, one gets $\delta \theta \ll 1^\circ $, 
%
hence, many (typically tens of) rocking steps $\delta_q$ are required to fully sample the diffracted intensity \cite{Pfeifer:2006aa,Ulvestad:2015aa,Robinson:2009aa}. 
Such a stringent condition is a direct consequence of the way a plane-wave extracts
the structural information about the sample.
More specifically, let us assume the usual Born  approximation within the scalar diffraction theory \cite{mertz2019}. In this case, the 3D spatial distribution of diffracting sources is given by $\psi=P\times \rho$ , with $P$ the beam and $\rho$ the complex-valued sample \cite{GuizarSicairos:2012aa,Rodenburg_2007,Godard:12}. The diffracted field at the detector plane, located in the far-field, is  a "slice" of the following 3D quantity \cite{Vartanyants_2001}: 
\begin{equation}
    \begin{array}{rcl}
    \Psi &:=& \mathcal{F} \psi\\ 
    &=& (\mathcal{F}P) \otimes  (\mathcal{F}\rho) 
    \label{eq:exit_field_focused_beam} 
    \end{array}
\end{equation}
where $\otimes$ denotes the 3D convolution operator and $\Fc$ is the 3D FT operator. 
In  standard BCDI, the illumination is a plane wave  
(i.e., $\mathcal{F} P$ is a centred Dirac distribution) and we deduce 
from \eqref{eq:exit_field_focused_beam} that the diffracted intensities 
in the BCDI data-set are actually drawn from the 3D intensity of the FT 
of $\rho$ itself. The sampling conditions for the BCDI 
experiment are then driven by the sample, and by the sample only.
When the illumination $P$ is spatially inhomogeneous (e.g., with the introduction of a MPP) the associated convolution  kernel $\mathcal{F}P$ spreads $\mathcal{F}\rho$ along a direction 
in reciprocal space that is perpendicular to the incident beam. In this situation, extracting a single slice in \eqref{eq:exit_field_focused_beam}   
actually samples $\mathcal{F}\rho$ \textit{concurrently} at various positions along the rocking curve scan.
The above reasoning suggests that the illumination modifies the range of information which is accessible from a single detector slice, and thus, may relax the oversampling requirements.
In Figure~\ref{fig:general_scheme}, we illustrate this idea, with three specific configurations corresponding to three standards of microscopy modalities that we discuss in the next paragraph. \\

\indent
(i) The first one is the \textit{focused beam} shown in panels $a$ and $a^\prime$ in Fig.~\ref{fig:general_scheme}, which is at the foundation of ptychography in CDI \cite{Rodenburg_2007,Thibault:2008aa,Hruszkewycz:2017ua,Chamard:2015aa}. For a finite illumination beam size, with transversal width $\Delta^{foc}$, the frequency mixing spans a window with a transversal extension of $\Delta \propto 1/\Delta^{foc}$ in reciprocal space. Along the RC direction, a single detector slice is then sensitive to the structural information within the domain $\sim \Delta \sin{\theta_{\tiny{B}}}$, as $\theta_{\tiny{B}}$ coincides with the incidence angle for a symmetric reflection.
Such an increase in sampling efficiency of the reciprocal space is obviously counterbalanced by the inability 
to extract this information over a large spatial field-of-view (FOV), due to the limited size of the focused probe. A  method providing a large FOV with relaxed sampling conditions in reciprocal space is thus,  desirable. (ii) The second configuration borrows its principle from \textit{structured illumination microscopy} (SIM), 
a super-resolution method developed in fluorescent (incoherent) microscopy in the early 2000's 
\cite{Gustafsson:2000aa}.   
To the best of our knowledge,  SIM has never been adapted to BCDI. Yet, the simplicity of the mechanism at work to extract frequency information otherwise inaccessible with an uniform illumination makes it particularly appealing 
\cite{Gustafsson:2000aa,Fiolka:2019aa,Idier:2018aa}.
In SIM, a pair of coherent sources located in the aperture plane 
generates an harmonic illumination that is invariant along the beam direction (see 
panel $b$ of Fig.~\ref{fig:general_scheme}) resulting in a highly inhomogeneous illumination over a large FOV in the transverse plane. 
Let us assume that the FT of $P$ is a pair of Dirac (pseudo-)functions located 
in $\pm \bar{\qb}$ with $\bar{\qb}$  a momentum transfer vector perpendicular 
to the incoming wave-vector $\mathbf{k}_i$ and  such that $2 |\bar{\qb}| = 
\Delta = 1/\Delta^{sim}$, where $\Delta^{sim}$ is the spatial period of the illumination (see panel $b^\prime$ of Fig.~\ref{fig:general_scheme}).
%
%
The diffracted wave field \eqref{eq:exit_field_focused_beam} is 
then a pair of shifted copies of the original Bragg peak:
\begin{equation}
    \Psi_{\text{SIM}}(\qb) = 
    (\mathcal{F}\rho)(\qb - \bar{\qb}) 
    + 
    \left(\mathcal{F}\rho\right) (\qb + \bar{\qb}).
    \label{eq:SIM_exit_field}
\end{equation}

The relation above indicates that the detector centred at the Bragg-peak 
provides
simultaneously the information that would be obtained from two successive measurements with
a plane-wave, shifting the peak along  $\widetilde{\kb}_3$ by $|\bar{\qb}| \sin \theta_B$ and
$-|\bar{\qb}| \sin \theta_B$, respectively; i.e., it performs a \emph{multiplexed} sampling of the Bragg peak.
%
This capacity to extract an information content otherwise inaccessible with an uniform illumination is at the very core of SIM  in fluorescence microscopy \cite{Gustafsson:2000aa,Fiolka:2019aa}.
This elegant and simple strategy has nevertheless two limitations.     
The first one relates to the use of pinholes to generate the pair of coherent x-ray sources at the aperture plane. This implies a vast loss of flux at the sample position, which is highly detrimental to the counting statistic of the data-set. 
The second limitation relates to the intensity of the harmonic pattern that vanishes periodically in the transverse plane. Clearly, because the sample cannot be retrieved at these regions that haven't received any photons, additional measurements with shifted positions of the illumination pattern are required.   
This issue is not specific to harmonic patterns as fully coherent, strongly inhomogeneous illuminations are probably difficult to achieve practically without vanishing intensities in some places. 
(iii) The third configuration depicted in panel $c$ in Fig.~\ref{fig:general_scheme} is a speckle illumination generated by a random phase-mask located in the aperture plane. 
We discuss below the design of such illuminations and their properties in the framework of BCDI acquisitions.   

\subsection{Speckle beam for BCDI acquisitions}
\label{sec:speckle_beam}

\begin{figure*}[htbp]
\begin{center}
\includegraphics[width=1. \linewidth]{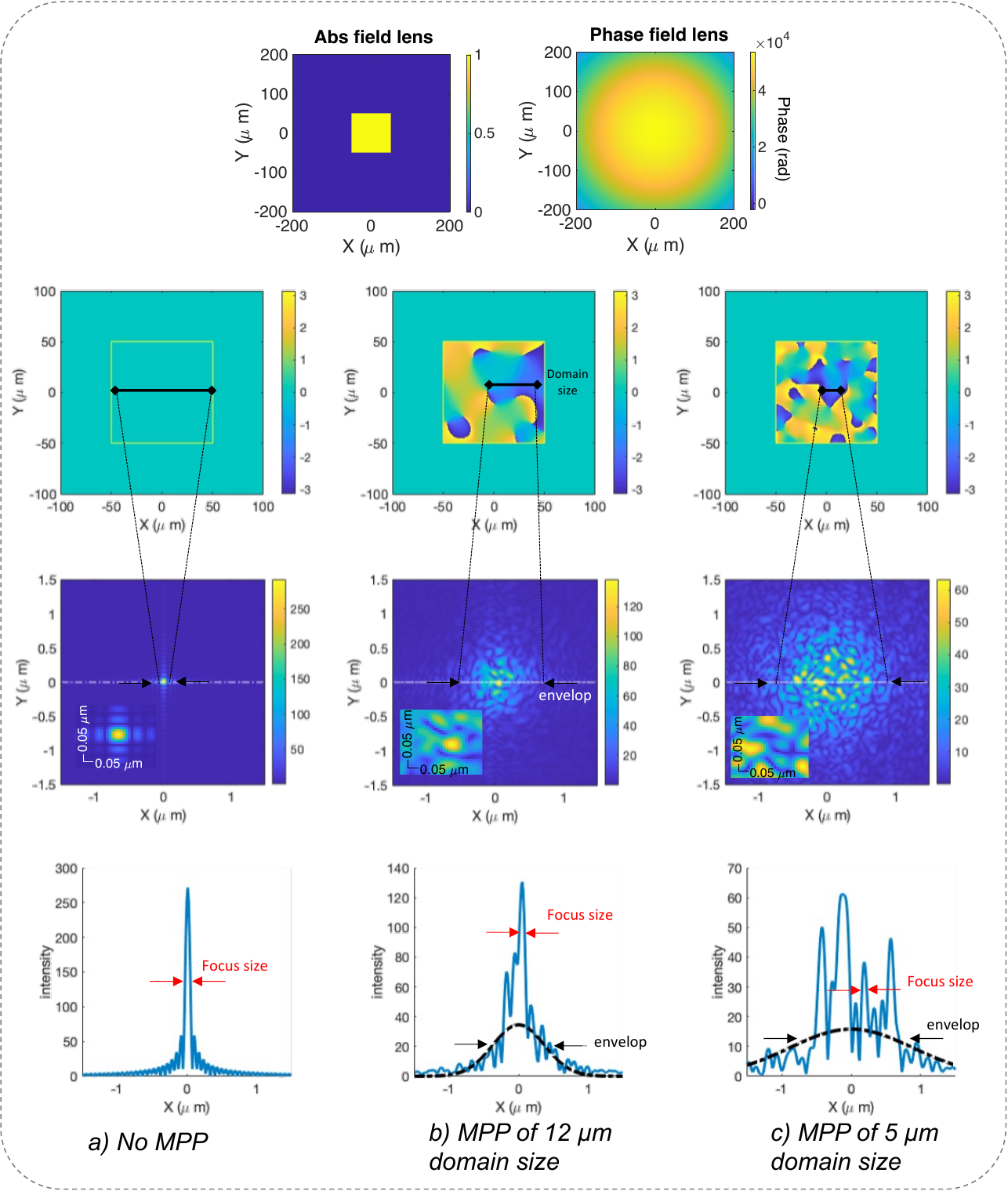}
\caption{Speckle beams achieved with the combination of a focusing lens producing a focal spot of 50 nm and two modulator phase plates (MPP) of different domain sizes: 5 and 12 $\mu$m. The first row represents the lens transfer function. The second row shows the phase configuration of each MPP. The third row displays the resulting illumination at the focal plane (where the sample is) and the insets are a zoom into the central part and provide a measure of the speckle size. Finally, the last row shows a 1D cut through the horizontal direction of the illumination. More details can be found in the Supplementary Information.}
\label{fig:MPP_spbeam}
\end{center}
\end{figure*}

\indent%
At hard x-rays energies, a speckle illumination can be achieved by manipulating the incoming beam wavefront with a phase plate, called \emph{modulator} (MPP). This MPP consists in a uniform optical aperture where phase-shifting domains are randomly distributed. When a coherent x-ray beam illuminates the MPP, the phase of its wavefront is strongly perturbed. The focusing of such a beam with an optical device like diffractive lenses (e.g. Fresnel zone plate\cite{Jefimovs:2007aa} or Multi-Laue lenses \cite{Kang:2006aa}) or refractive/reflective lenses (e.g. compound refractive lenses \cite{Snigirev:1996uz} or KB mirrors \cite{Kirkpatrick:48}), produces intense interferences at the focal plane resulting in a \emph{speckled beam} (see section 1 of Supplementary Information for more details).
%
As shown in Fig.~\ref {fig:MPP_spbeam}, the features of the speckled beam are fully controlled by the MPP and by the lens. 
In particular, the  size of the speckle grain is determined by the lens aperture; it is then identical to the lateral extension of the focal spot that would be obtained with the focusing optics alone. The FOV of the speckled beam (i.e., the FWHM of the speckle envelope) is governed by the correlation-length of the random phase in the MPP; it is then directly related to the size of the phase domains in the modulator (Fig.~\ref{fig:general_scheme}$a, a$'). 

\indent%
Similarly to SIM, the intensity of the generated speckle field is going to vanish in some lateral positions. To minimize the impact of these dark regions, a few RCs with the speckle illumination shifted laterally are needed to retrieve a robust estimate of the sample structure.
However, the isotropic nature of the correlation of the speckle provides more flexibility than the harmonic pattern to achieve a non zero total photon deposit. Standard piezoelectric translation stages commonly used in several synchrotron beamlines can routinely achieve shifts comparable to the speckle size (e.g., $\sim$20-200 nm) \cite{Mastropietro:2017un,Chamard:2015aa}. 

\indent%
MPP illumination is therefore a suitable strategy for structured illumination experiments, for its ability to generate a speckle illumination over a large FOV and with almost no loss of coherent flux (besides the absorption by the material of the phase modulator).
Moreover, similarly to the harmonic illumination or the focused beam, the speckled beam also performs the "frequency-mixing" action, which is expected to foster the sampling efficiency along the RC direction (Fig.~\ref{fig:general_scheme}$c$').
In the next section, we derive the relaxed sampling conditions along the RC that can be achieved with specific speckle beam parameters.

\subsection{Relaxed sampling condition along the RC}
\label{subsubsec:spBCDI_efficient_sampling}

\begin{figure}[htbp]
\begin{center}
\includegraphics[width=1.0 \columnwidth]{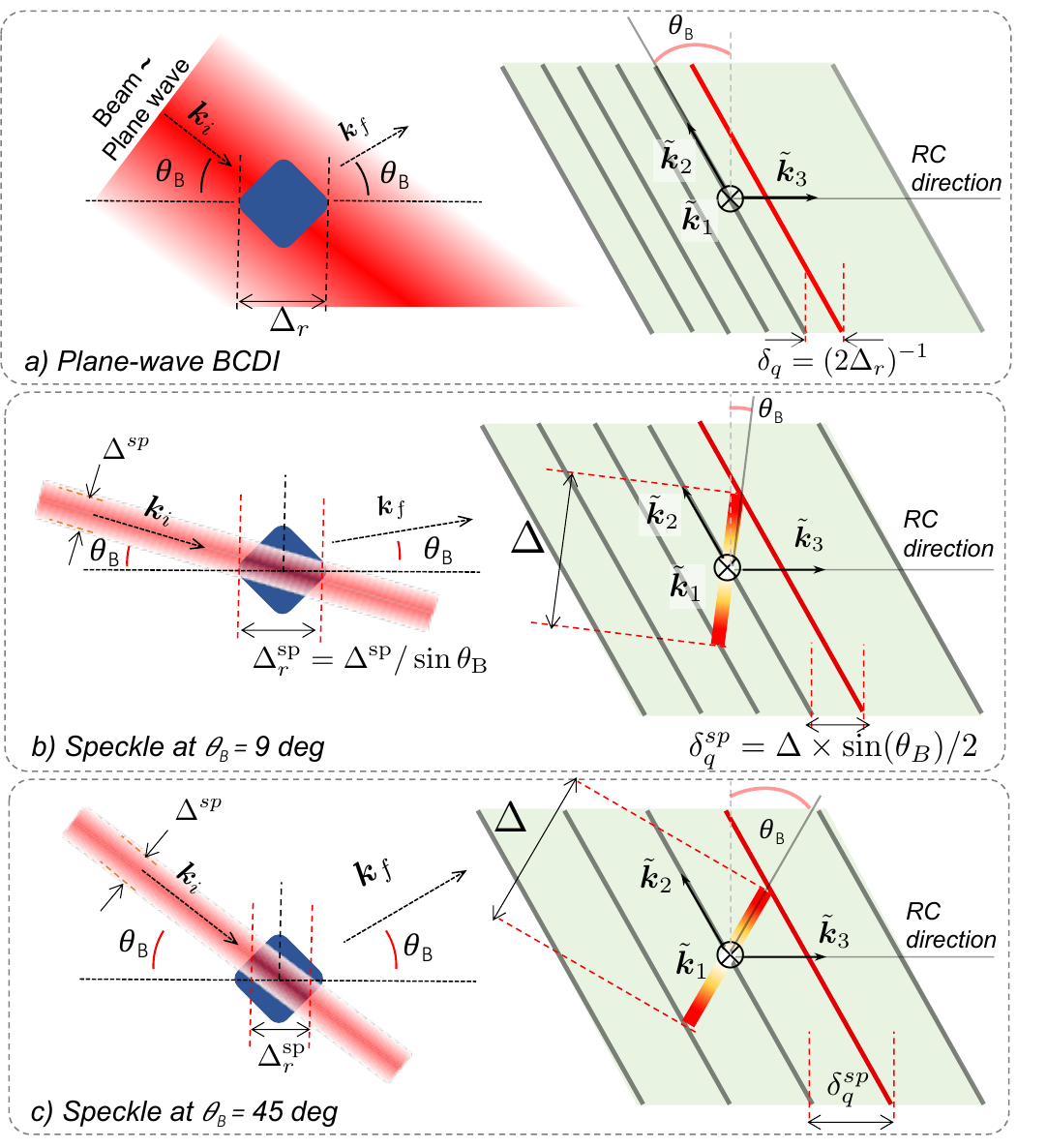}
\caption{Effect of the incident angle in the OSR sampling of reciprocal space and comparison with standard OSR conditions for standard (plane wave) BCDI .  }
\label{fig:thetaB_OSR}
\end{center}
\end{figure}

\indent%
Speckle-BCDI exploits the multiplication of shifted copies of the object FT arising from a non-uniform illumination to increase the amount of 3D information in each 2D slice of the RC scan (see Eq. \ref{eq:SIM_exit_field} and Fig. \ref{fig:general_scheme}). This effect is predicated on the structural information distribution in reciprocal space resulting from the convolution between the illumination and the object FTs. Thus, the FT of the illumination determines the extent of the reciprocal space that a single detector slice is capable to probe. This sets a lower limit for the reciprocal space step along the RC direction, $\delta^{sp}_q$, which directly leads to the minimum angular sampling step along the RC scan. To ensure the possibility to retrieve the phase from the intensity data-set, the Nyquist-Shannon criterion should be verified for the speckle size along the RC direction, \emph{i.e.}, the speckle footprint $\Delta^{sp}_r$ expressed  as follows (see Fig. \ref{fig:thetaB_OSR}): 
\begin{equation}
\Delta^{sp}_r = \frac{\Delta^{sp}}{\sin{\theta_B}}. 
\end{equation}
where $\Delta^{sp}$ is the lateral extension of the speckle grain. It results that,
\begin{equation}
\delta^{sp}_q = (2\Delta^{sp}_r)^{-1} = \frac{\sin{\theta_B}}{2\Delta^{sp}} = \frac{\Delta \sin(\theta_B)}{2}
\label{eq:Delta_sp_q}
\end{equation}
Substituting Eq.~\ref{eq:Delta_sp_q} in Eq.~\ref{eq:OSR_stBCDI_recipspace}, an estimate of the minimum sampling ratio $\sigma^{min}_{RC}$ is obtained:
\begin{equation}
\sigma^{min}_{RC} = \frac{1}{\delta^{sp}_q \Delta_r}  = \frac{2\Delta^{sp}}{\Delta_r\sin{\theta_B}}
\label{eq:OSR_min}
\end{equation}
It defines the  minimum sampling ratio along the RC scan, for which the inversion of the intensity data is still expected to be unique and robust.

This rule is illustrated in Fig. \ref{fig:thetaB_OSR}, where is included the sampling conditions for standard BCDI, \emph{i.e.}, $\sigma_{RC} = 2$ (panel $a$). It also displays the impact of the incidence angle in the allowed separation between detector slices. The interest of this approach is further estimated by considering the following experimental example, where a cubic sample of edge $L = 700$ nm and diagonal $\Delta_r = 990$ nm is illuminated at two different incident angles, 9$^{\circ}$ and 45$^{\circ}$, by three different focused beams, characterized by their speckle sizes, 50, 80 and 100 nm. Table~\ref{tab:beam_parameters_factor2} summarizes the resulting $\sigma^{min}_{RC}$ values, which are further compared to the case $\sigma_{RC} = 2$, the minimum sampling criterion for the plane-wave BCDI approach. Important gains in sampling efficiency, of about one order of magnitude, can be expected at larger incidence angles and smaller speckle sizes. A refined estimation of the accessible gains will be further provided with the numerical experiments summarized in the next section.
\begin{table}[h]
\caption{Illustration of the gain in sampling induced by the use of a speckle beam compared to the plane wave illumination at the Shannon-Nyquist criterion ($\sigma_{RC}=2$), for different incident angles and different beam speckle sizes. The sample is a cube of edge $L = 700$ nm and diagonal $\Delta_r = 990$ nm. Symmetrical reflection is considered so that the incident angle is equal to the Bragg angle.}
\newcolumntype{g}{>{\columncolor{Gray}}c}
\begin{center}
\begin{tabular}{|cl||g|c||g|c||g|c|}
\hline
$\Delta^{sp}$ &[nm]  			& \multicolumn{2}{c||}{50}   &\multicolumn{2}{c||}{80}	&	\multicolumn{2}{c|}{100}	\\
\hline
$\theta_B$ &[degree] &9	  &	45	    &9	   &	45	&	9	&	45	\\\hline
$\sigma^{min}_{RC}$ & 				    &0.64 &	0.14	&1.0   &	0.23&	1.29&	0.29\\\hline
Gain wrt $\sigma_{RC} = 2 $	& &3.1  &	14	    &1.9   &	8.8	&	1.5	&	7   \\
\hline 
\end{tabular}
\end{center}
\label{tab:beam_parameters_factor2}
\end{table}

\section{Numerical demonstration of spBCDI}
\label{sec:numerical_simulations}

\begin{figure*}[htbp]
\begin{center}
\includegraphics[width=1. \linewidth]{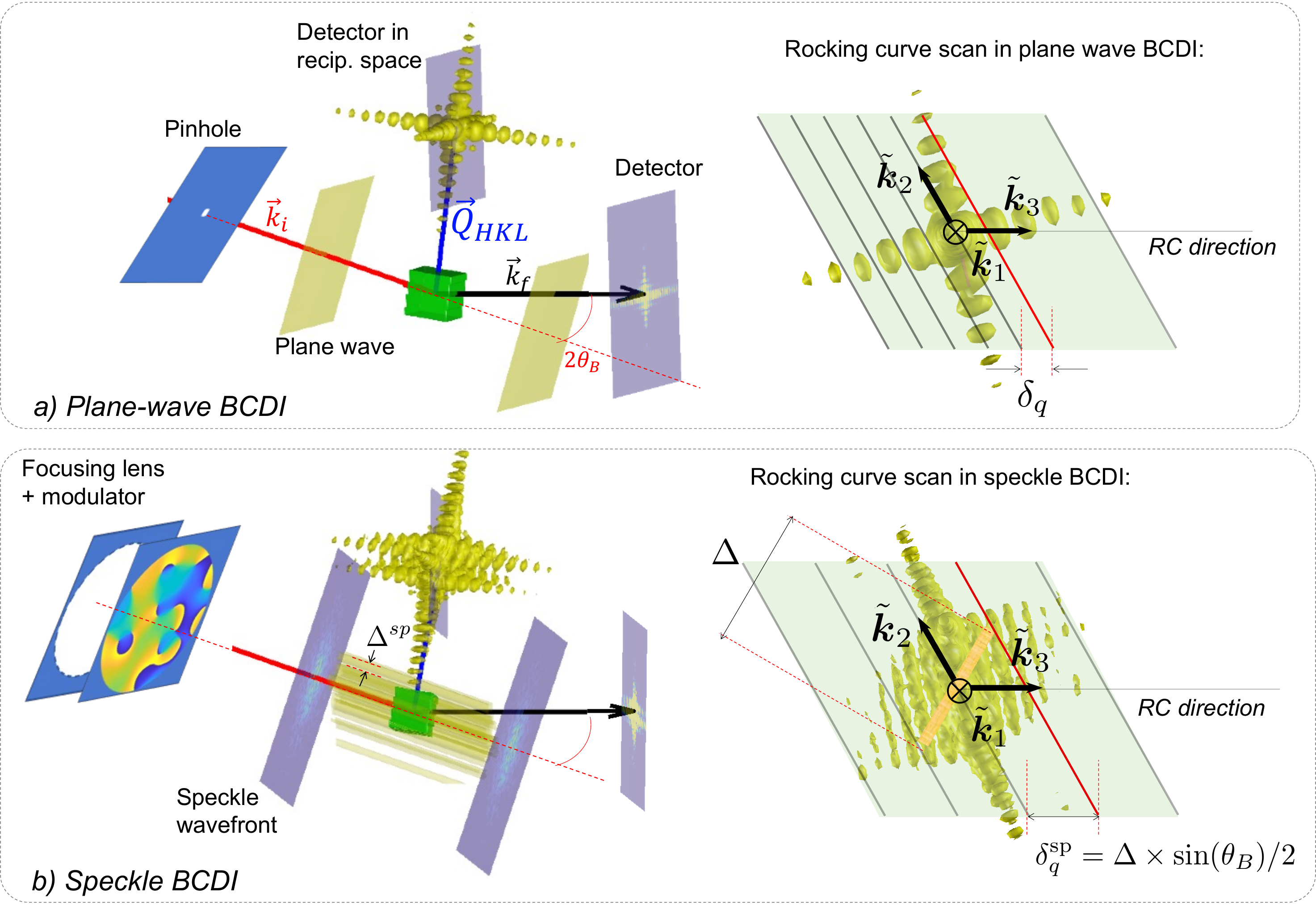}
\caption{Schematic description of the standard (plane-wave) BCDI (a) \textit{versus} speckle BCDI (b).}
\label{fig:pw_vs_sp_BCDI}
\end{center}
\end{figure*}

Fig. \ref{fig:pw_vs_sp_BCDI} illustrates the principle of spBCDI versus standard 
(i.e., plane-wave based) BCDI  on a perfect cubic crystal. In standard BCDI, the 
cube is fully illuminated by a plane wave, producing in the far field the typical sinc function associated to a cube FT. The fringes emanating from the Bragg peak are perpendicular to the cube facets and their width is $2\pi/\Delta_r$ in each direction.  Conversely, in spBCDI the cube is illuminated by a strongly non-uniform speckle pattern, producing in the far field a more complicated diffraction pattern. 
%
As explained in Sec.~\ref{sec:speckle_beam}, the resulting diffracted field  is the convolution between the FT of the speckle illumination and the FT of the object (see Eq.~\ref{eq:exit_field_focused_beam}). The diffracted 3D intensity results in a series of copies of the Bragg peak shifted (and weighted) continuously in the plane transversal to the beam propagation. This process replicates the information carried by the FT of the object along a direction, which aligns partially  with the RC direction. In this section dedicated to the numerical demonstration of spBCDI, we first describe the design of the numerical model including a 3D complex-valued sample and a highly structured speckle beam. Results and performances of the newly developed iterative inversion scheme are presented in a second part.  

\subsection{Speckle-BCDI simulated measurements}

\indent%
The numerical experiments highlight the interest of spBCDI for efficiently imaging 3D crystalline samples. We anticipate that this new method yields different performances as a function of a large set of parameters, such as the RC sampling frequency, the accessible intensity dynamical range, the incidence angle, the crystalline strain complexity, the availability of the sample shape knowledge, the speckle distribution, to cite only a few. In this work, we focus our investigations on a selection of them: the RC sampling frequency, the signal-to-noise (SNR) ratio, the sample support and the incidence angle. The effect of the other parameters will be discussed in the last section.\\ 

\indent
Our numerical tests involve a highly distorted crystal, produced from the cubic crystal depicted in Fig.~\ref{fig:pw_vs_sp_BCDI}, multiplied by a complex-valued field $\exp{i \phi}$. In BCDI, this phase $\phi$ is mathematically described by $\phi = \mathbf{G}_{HKL}\cdot\mathbf{u}$, $\mathbf{u}$ being the atomic displacement vector arising from the internal distortion of the crystalline lattice \cite{Pfeifer:2006aa,Robinson:2009aa,Newton:2020aa}. For further comparison with previously published works, the strain and its distribution are further described in the section 2 of Supplementary Information. Given the real space sample size and the phase distribution, the non-homogeneous part of the strain is obtained assuming that $\mathbf{G}_{HKL}$ is the 104 Bragg reflection of calcite \cite{Clark:2015aa,Geissbuhler:2004aa}. The simulated displacement field extends over one order of magnitude, with a large part of the strain being in the order of a few $10^{-4}$ and a significant amount of the strain increasing up to a few $\pm 4\cdot10^{-3}$, spanning a total non-homogeneous strain distribution of almost $10^{-2}$. The relevance of the strain field in comparison with experimental studies is discussed in the last section of the paper.   \\

\begin{figure*}[htbp]
\begin{center}
\includegraphics[scale=.5]{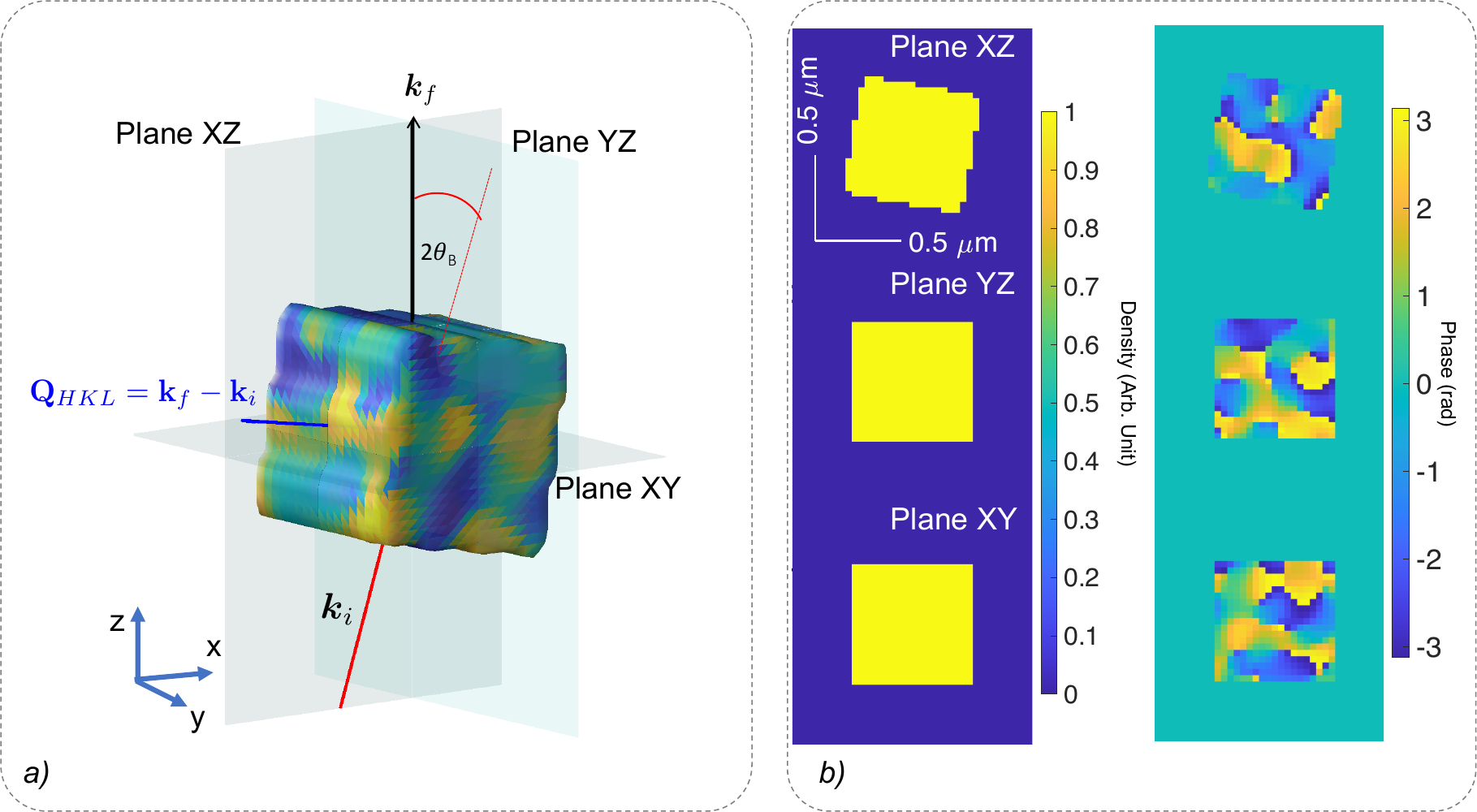}
\caption{(a) 3D representation of the non-homogeneously strained crystalline cube, in the scattering geometry corresponding to an incident angle of 9$^\circ$.  The color map encodes the phase at the surface. The vectors represent the incident $\kb_i$ and exit $\kb_f$ beams and $\Qb_{HKL}$ is the momentum transfer matching the $HKL$ Bragg condition. (b) Cuts through different planes of the modulus (left) and the phase (right) of the cubic crystal}
\label{fig:cube3D}
\end{center}
\end{figure*}

A 3D image of the system with cuts of its modulus and phase through the main orthogonal planes (\textit{i.e.} $XY$, $XZ$, $YZ$) is represented in Fig. \ref{fig:cube3D}. The numerical window of 64 x 64 x 90 pixels (with isotropic pixel sizes of 30 nm) has been chosen such that the cube of $\sim$ 20$^3$ total pixel span fits at least three times in each direction (the pixels outside of the object support have been settled to zero). This ensures that the autocorrelation function  of $\rho$ is fully contained in the simulation array, and hence, prevents cycling aliasing \cite{Miao:1998aa}. Equivalently, it guarantees that the diffracted intensity is oversampled with sufficient rates of $\sigma_x=3.2$, $\sigma_y=3.2$ and $\sigma_{RC}=4$.
\\
\indent The speckle illumination is simulated considering a focus size of 50 nm and a MPP producing an envelop of FWHM $\sim$ 2 $\mu$m, which fully illuminates the cubic crystal of 700 nm edge (panel $b$ of Fig. \ref{fig:MPP_spbeam}). In this study, two extreme incident angles are chosen. The shallow one is set at 9$^\circ$, which corresponds to the Bragg angle for the (104) calcite Bragg reflection at a beam energy of 13 keV. Later on, a 45$^\circ$ incident angle is chosen to test the limit of the method. In this case, the real space pixel size along the RC direction is reduced by a factor of three (\emph{i.e.} $\sim$ 10 nm), to allow for a larger number of diffraction patterns along the RC, without changing the sampling frequency in that direction. In this way, extreme undersampling conditions can be investigated.

\indent%
The above described sample and beam determine the distribution of the over-sampled 3D intensity data-set, which is given by the square modulus of the scattering volume FT, defined as the 3D product of the beam and sample. To  test the ability of the phase-retrieval algorithm to reconstruct under-sampled spBCDI data (\emph{i.e.}, data acquired with larger angular steps) we build a set of 3D binary masks discarding, in the original oversampled 3D intensities, the appropriate number of slices along the RC direction. Each 3D mask corresponds to a specific sampling condition along the RC direction and is  used within the reconstruction algorithm to select the suite of planes to invert at a particular sampling condition. This method is then a flexible way to simulate various sampling ratios $\sigma_{RC}$. In our simulation, $\sigma_{RC}$ varies from 4 down to 0.06. Hence a ratio of 4 corresponds to the complete data-set, while a ratio of 2 corresponds to sampling at the Shannon-Nyquist frequency and a ratio of 0.5 corresponds to measuring 1 over 8 diffraction patterns along the RC direction. Moreover, because a single speckle illumination needs to be shifted to prevent any zero photon deposit in the sample volume, each spBCDI data-set consists of a series of four RC scans at shifted positions of the beam and, thus, increasing proportionally the number of diffraction patterns. \\
Finally, different SNRs have been introduced to push the test to the limits of the method in the context of high rate measurements. To reproduce realistic SNRs, the signal intensity has been chosen to range between 2000 down to 35 ph/pixel at the maximum of the Bragg peak (before the binary mask is applied). For this, the 3D diffracted data-sets are normalized to the SNR levels and Poisson noise is eventually added to simulate stochastic fluctuations in the photon count rate. This provides a convenient way of characterizing the noise level and it is applied to both BCDI and spBCDI data-sets. 
Note however that with speckle illuminations, there are multiple copies of the Bragg peak in the scanned volume. Therefore, we observed that the
total number of diffracted photons might be in some cases two times larger than for plane wave BCDI
where there is only one Bragg peak, for the same SNR value.\\
A comparison of spBCDI performances with BCDI is included. The plane-wave BCDI simulation assumes a full planar illumination of the sample. The 3D diffraction pattern is produced for $\sigma_{RC}= 4$, and the binary mask is applied to adjust the (effective) sampling step $\delta_q$ along RC. Noise is added to the data-set at a SNR level of 2000. Its inversion was achieved with a standard combination of error reduction (ER) and hybrid input-output (HIO) inversion algorithms, as described in the literature \cite{Fienup:1982aa,Marchesini:2007aa}. Table~\ref{tab:DP_number} summarizes the main characteristics of the presented results. It includes the total number of diffraction patterns forming each data-set and the gain in acquisition time, when successful reconstructions are achieved.\\
 
\begin{table*}[htbp]
\caption{Number of diffraction patterns (DPs) forming each data-set as a function of $\sigma_{RC}$ for BCDI and spBCDI. To prevent for any zero photon deposit in the sample volume for spBCDI, the RC scans are repeated four times, at  shifted beam positions (the beam shift is of $\sim$ 1.2 and 1.5 times the speckle size). A decrease in $\sigma_{RC}$ corresponds to a larger step along the RC direction. The numerical tests are performed for different SNR values, ranging from 2000 counts down to 35 counts. Whenever the reconstruction is successful, the gain is also indicated. Its quantification includes the number of diffraction patterns and the acquisition time. The reference gain, set to one, is taken for the plane-wave BCDI at $\sigma_{RC} = 2$ and SNR = 2000 counts. For the other experiments, the gain expresses as $(45/$DPs$) \times (2000/$SNR$)$. On each line, the best achieved gain  is highlighted. Non-reliable reconstructions (NR) or not performed tests ($-$) are also indicated.} 
 \begin{center}
\begin{tabular}{|c|r|c|c|c|c|c|c|c|c|c|}
\hline
\backslashbox{Experiment}{$\sigma_{RC}$}&&4    &	2   &	1.0	&	0.7   &   0.5      &   0.4    & 0.3   & 0.1      & 0.06 \\\hline\hline
BCDI, 1 RC      			    &	DPs & 90   &	45	&	23	&	16    &   11       &   9      &   7   & --       & -- \\
2000 counts     			    & Gain	&  0.5 &\textbf{1}&NR&	NR    &  NR        &  NR      &  NR   &          &  \\\hline
BCDI, 1 RC      			    &	DPs & 90   &	45	&	--	&	--    &   --       &   --     &   --   & --       & -- \\
1000 counts     			    & Gain	&\textbf{1}& NR &       &	      &            &          &        &          &  \\\hline \hline
spBCDI ($9^\circ$), 4RCs        & DPs   & 360  &   180  &	80	&	64	  &	  44       &   36     &  28   & --       & -- 	\\
500 counts                      & Gain  & 0.5  &   	1   &	2   &	2.8   & \textbf{4} &   NR     &  NR   &          &     	\\\hline
spBCDI ($9^\circ$), 4RCs        & DPs   & 360  &    180	&	80	&	64	  &	  44       &   36     &  28   & --       & -- 	\\
200 counts                      & Gain  & 1.2  &   2.5  & 5.6   &	7	  &	   10      &\textbf{12}&  NR  &          &  	\\\hline
spBCDI (at $45^\circ$), 4RCs    & DPs   & --   &    --	&	--	&	--	  &	   200     &   --      &  118  & 40      & 24    \\
200 counts                      & Gain  &      &   	    &	    &		  &	    2      &          &  3.8  &\textbf{11}& NR 	\\\hline
spBCDI (at $45^\circ$), 4RCs    & DPs   & --   &    --	&	--	&	--	  &	  200      &      --   &   --  & 40        & --       \\
100 counts                      & Gain  &      &   	    &	    &		  &	    4      &           &      &\textbf{22} &  	\\\hline
spBCDI (at $45^\circ$), 4RCs    & DPs   & --   &    --	&	--	&	--	  &	  200      &     --    &   --   & 40        & --     \\
70 counts                       & Gain  &      &   	    &	    &		  &	    5.7    &           &      &\textbf{32}  &  	\\\hline
spBCDI (at $45^\circ$), 4RCs    & DPs   & --   &    --	&	--	&	--	  &	  200      &       --  & 118     & 40        & --    \\
35 counts                       & Gain  &      &   	    &	    &		  &	    11     &           &    \textbf{22}  &    NR       &      	\\\hline

\end{tabular}
\end{center}
\label{tab:DP_number}
\end{table*}%

\begin{figure*}[htbp]
\begin{center}
\includegraphics[width = 1. \textwidth]{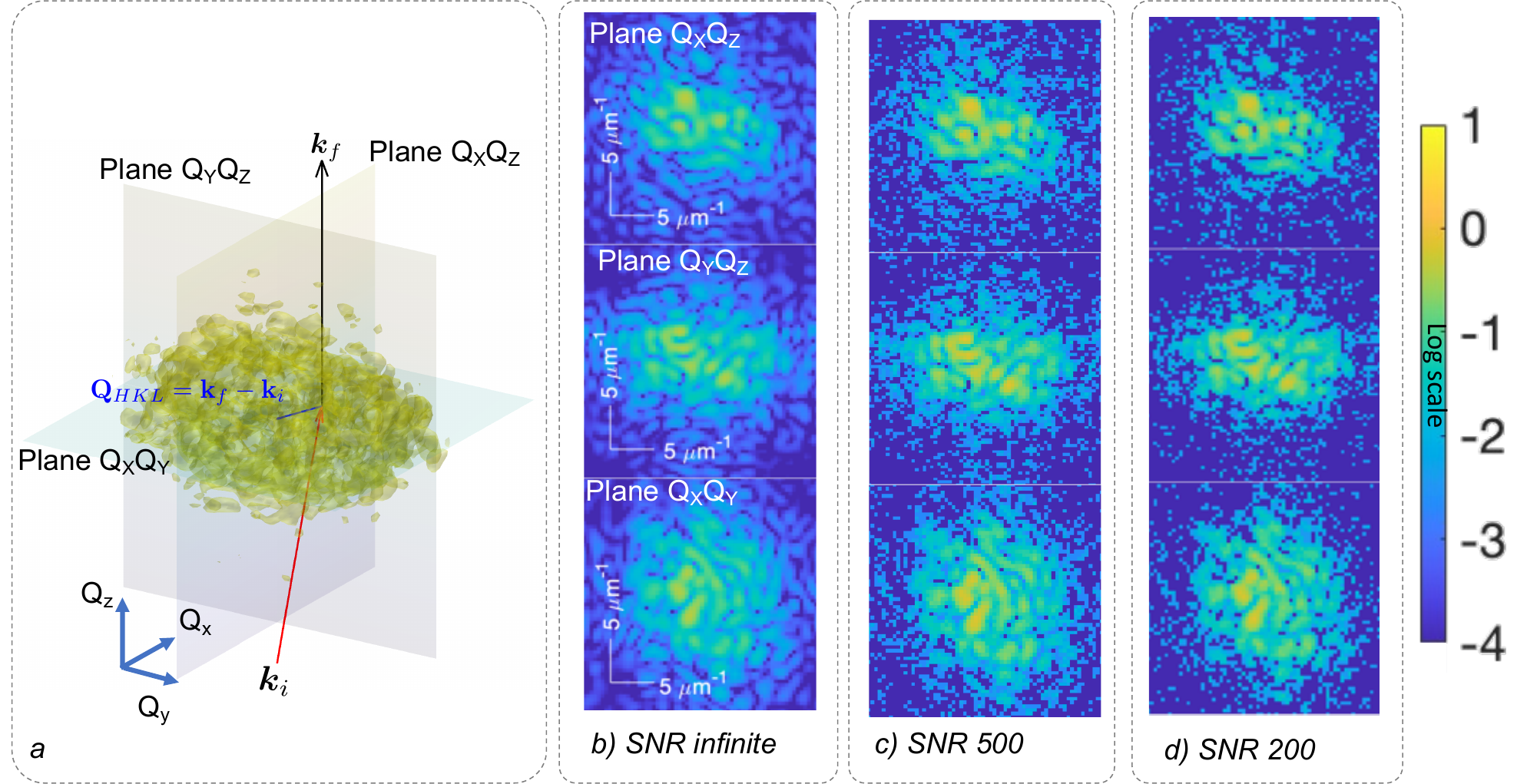}
\caption{(a) Normalized 3D diffracted intensity distribution around the $HKL$ Bragg peak (located at the head of the momentum transfer $\mathbf{Q}_{HKL}$, resulting from the combination of the incoming $\kb_i$ and outgoing $\kb_f$ wavevectors) and the three main planes of the reciprocal space. (b-d) 2D cuts of the  distribution of intensity in the planes highlighted in panel $a$ at infinite SNR first, and then at the two SNRs under study: 500 and 200, respectively. The detector planes corresponds to the $(Q_X,Q_Y)$ plane.}
\label{fig:SNR}
\end{center}
\end{figure*}

 The effect of finite SNR in the diffracted intensity is summarized in Fig.~\ref{fig:SNR}. It presents the 3D intensity distribution around the Bragg peak (Fig.~\ref{fig:SNR}$a$), located at $\mathbf{G}_{HKL}$, within an orthogonal reference frame ($\mathbf{Q_x},\mathbf{Q_y},\mathbf{Q_z}$) which differs from the tilted detection space  ($\mathbf{\tilde{k}_1},\mathbf{\tilde{k}_2},\mathbf{\tilde{k}_3}$) used in Figs. \ref{fig:BCDI_geometry} - \ref{fig:pw_vs_sp_BCDI}, because it is the conjugated version of the set of vectors $(\mathbf{x},\mathbf{y},\mathbf{z})$ in which the 3D object is represented in Fig. \ref{fig:cube3D}.
 Panels $b$, $c$ and $d$ display the normalized intensity distribution in the three  main planes of the reciprocal space, first in the noiseless case (SNR infinite) and then at SNRs of 500 and 200 counts, respectively.


\subsection{Phase retrieval strategy in speckle-BCDI}
\label{subsec:phase_retrieval}

\indent%
The inversion of the spBCDI simulated measurements was performed with a straightforward adaptation of the traditional \emph{error reduction} (ER) algorithm \cite{Fienup:1982aa}. More specifically,
we define the following criterion to minimize 
\begin{equation}
    J(\rho) = \sum_{p=1}^P \sum_{n=1}^N \Omega(\qb_n ; \sigma_{RC}) \times \left(\sqrt{\widehat{I}_p(\qb_n)} - \sqrt{\bar{I}_p(\qb_n ; \rho)}\right)^2.
    \label{eq:Criterion}
\end{equation}
 In this relation, $\widehat{I}_p(\qb_n)$ is the 3D experimental intensity  generated from the RC using the $p$-th position of the speckled beam at each pixel $\qb_n$ of the sampled volume; $\bar{I}_p(\qb_n; \rho)$ is the noise-free intensity expected for this measurement when the sample $\rho$ is given, see Eq.~\ref{eq:exit_field_focused_beam}. For a given sampling ratio along the RC, $\Omega(\qb_n;\sigma_{RC})$ is the 3D binary mask selecting the appropriate set of diffraction patterns within the RC that are considered to run the  algorithm.  
With this error metric, a gradient with respect to $\rho$ is derived and used in an iterative minimization algorithm to estimate the sample \cite{NoceWrig06}. The retrieved object corresponds to a  local minimum of the error metric and the algorithm stops when the experimental and calculated diffraction patterns are congruent within the noise level.\\

\indent%
We first investigate the reliability of the reconstructions when the  object shape is perfectly known; in this case, only ER iterations as described above are sufficient to compute the modulus and phase estimates for each pixel within the known support. When the support is unknown, we need to periodically update the support \textit{via} the shrink-wrap (SW) method \cite{Marchesini:2003aa}; this case will be  considered later in this section. In all cases we assume that the speckle illumination (in modulus and phase), and its exact position for each RC scan are perfectly known.   

\begin{figure*}[htbp]
\begin{center}
\includegraphics[width=1.\textwidth]{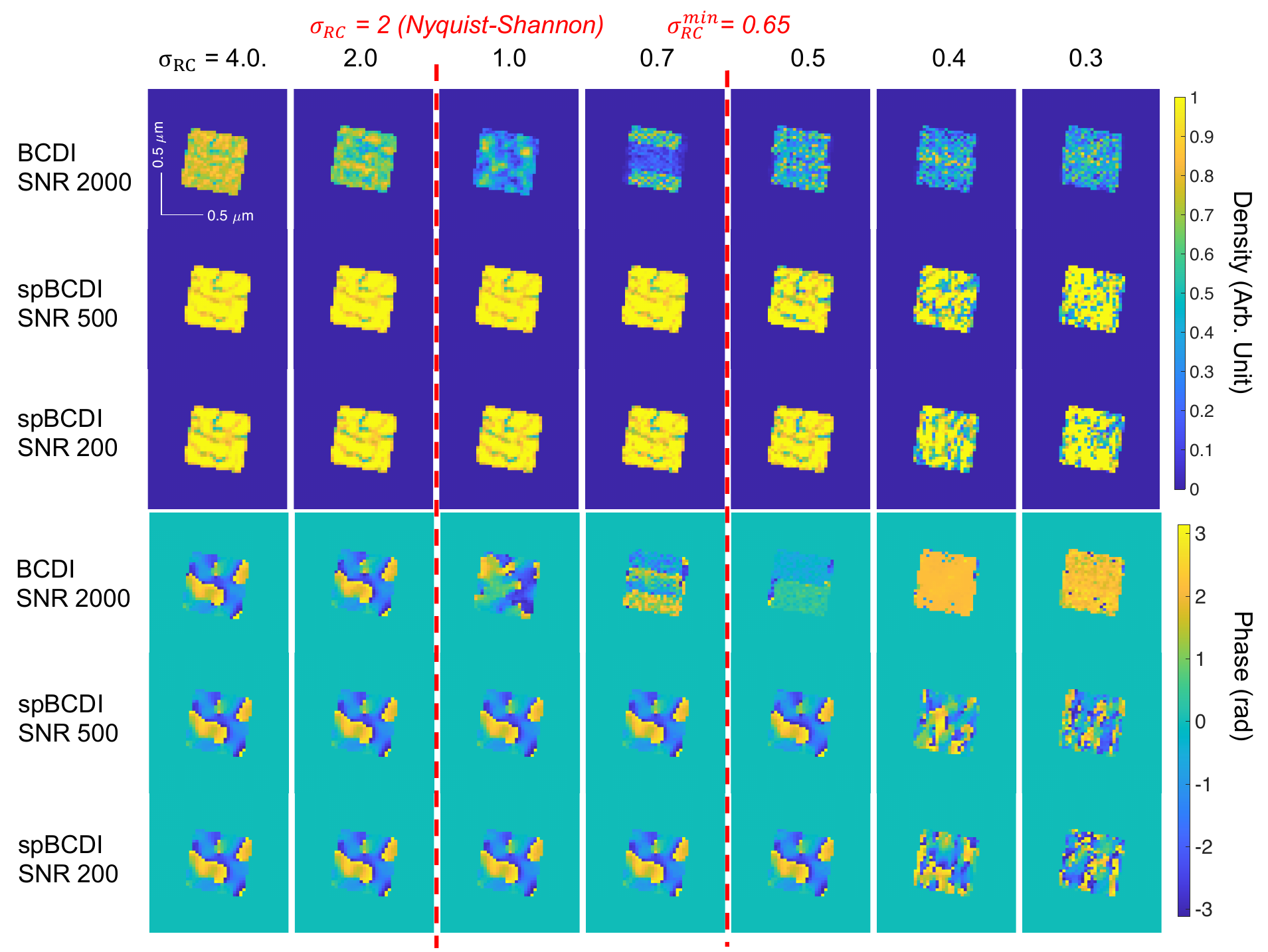}
\caption{Speckle BCDI (spBCDI) \textit{versus} standard plane-wave BCDI: A comparison study for $\theta_B = 9^{\circ}$. Reconstruction results are shown for different sampling ratio $\sigma_{RC}$ and different SNR levels. To avoid regions of zero photon deposit for speckle illumination, each spBCDI data sets consists of a series of four RC scans taken at shifted positions of the beam. These tests assume known support for the sample.}
\label{fig:resuls_9deg_OSR}
\end{center}
\end{figure*}

\begin{figure*}[htbp]
\begin{center}
\includegraphics[scale=.27]{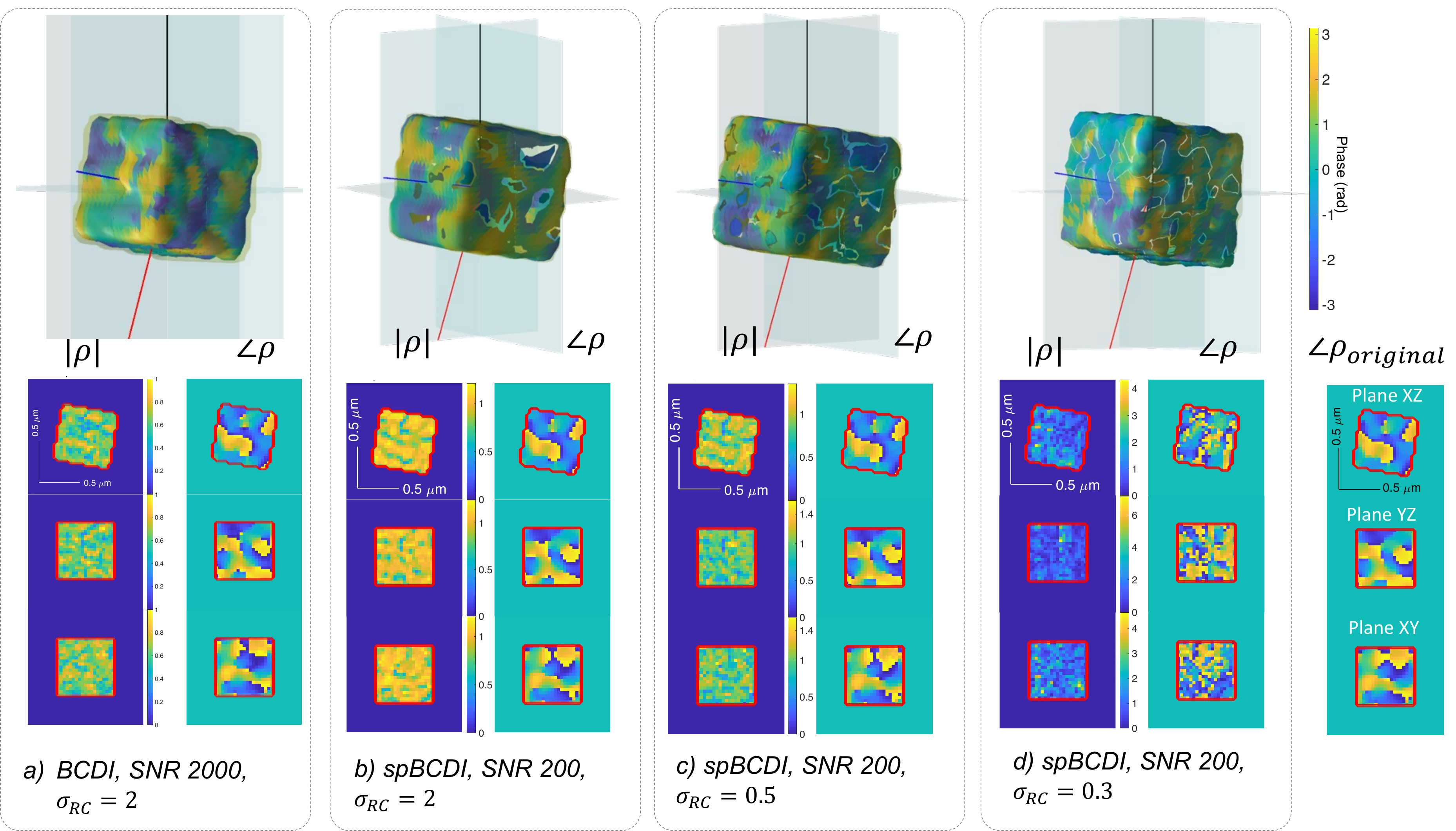}
\caption{3D reconstructions of BCDI (panel $a$) and spBCDI (panels $b$ - $d$) data sets when the support is exactly known, for $\theta_B = 9^{\circ}$. The modulus, $|\rho|$, and the phase, $\angle \rho$, of the object are shown while the contour of the support is indicated in red. This latter is also shown in the 3D pictures with a semi-transparent yellow surface. The red, black and blue arrows indicate respectively $\kb_i$, $\kb_f$ and $\Qb_{HKL}$. Each column from $(a)$ to $(d)$ contains a label, which specifies the RC sampling frequency, $\sigma_{RC}$, and the SNR values.}
\label{fig:resuls_3D_9deg_OSRs_exact_support}
\end{center}
\end{figure*}

\indent%
\paragraph{Performances with respect to plane-wave BCDI:}
To start with, we compare the performances of spBCDI with respect to plane-wave BCDI. The comparison is done at a shallow incidence angle of 9$^\circ$, a configuration for which the decrease in $\sigma_{RC}^{min}$ is expected to be limited (see Eq. \ref{eq:OSR_min}). 
Figure~\ref{fig:resuls_9deg_OSR} summarizes the reconstruction tests performed for a series of data-sets consisting in four spBCDI RC scans at shifted beam positions. In these simulations, $\sigma_{RC}$ varies from the initial value of 4 down to 0.3 and SNR values are set to 500 and 200 counts, respectively. We further assume that the support of the object is exactly known, \emph{i.e.} the most favorable condition to set a benchmark. To display the results in a single figure, only the reconstructed moduli and phases in the scattering plane ($XZ$) are presented. Note that for each case, multiple reconstructions were performed to investigate the robustness of the inversion and the relevance of the obtained results. \\
While BCDI starts failing for $\sigma_{RC} \leq 2$ at a SNR of 2000 counts and even for $\sigma_{RC} \leq 4$ for an SNR of 1000 counts (see Table \ref{tab:DP_number}), the spBCDI approach still provides reliable reconstructions for a $\sigma_{RC}$ value down to 0.5, for the two investigated SNRs. A more detailed description regarding the reconstruction quality, is shown in  Fig.~\ref{fig:resuls_3D_9deg_OSRs_exact_support}, which presents the retrieved moduli and phases for three different planes, in 4 selected cases: standard BCDI at $\sigma_{RC}$ = 2 and SNR of 2000, and three spBCDI reconstructions at SNR of 200 for $\sigma_{RC}$ values of 2, 0.5 and 0.3. It demonstrates the  great fidelity achieved in spBCDI reconstructions, specially regarding the retrieved   phase, at a very low $\sigma_{RC}$ of 0.5, which is well below the standard value of two established by the Shannon-Nyquist theorem and slightly below the expected theoretical value of 0.65. As often observed in presence of noise in the intensity patterns, the reconstruction of the modulus is more challenging at finite SNR \cite{Godard:12}. However, despite of degraded modulus reconstruction, inversion of spBCDI data-sets successfully retrieves the phase in heavily distorted crystals and even with a  reduced number of diffraction patterns (Table \ref{tab:DP_number}). An empirical limiting value of $\sigma_{RC} = 0.4$ is found, which corresponds to the limit below which the algorithm fails systematically to reconstruct neither the modulus nor the phase (see panel $d$ of Fig. \ref{fig:resuls_3D_9deg_OSRs_exact_support} where the reconstruction at a $\sigma_{RC} = 0.3$ is presented). In this specific case, accounting for the 4 RC scans per data-set in spBCDI, a $\sigma_{RC}$ of 0.4 and a SNR of 200 yields a measurement  time of a factor 12 shorter than for the standard plane-wave BCDI data-set simulated at $\sigma_{RC} = 4$ \emph{i.e.} the minimum $\sigma_{RC}$ value needed to provide a reasonable estimate of the distorted phase field for an SNR of 2000. This gain represents already a significant reduction of the acquisition time.\\

\begin{figure*}[htbp]
\begin{center}
\includegraphics[scale=.27]{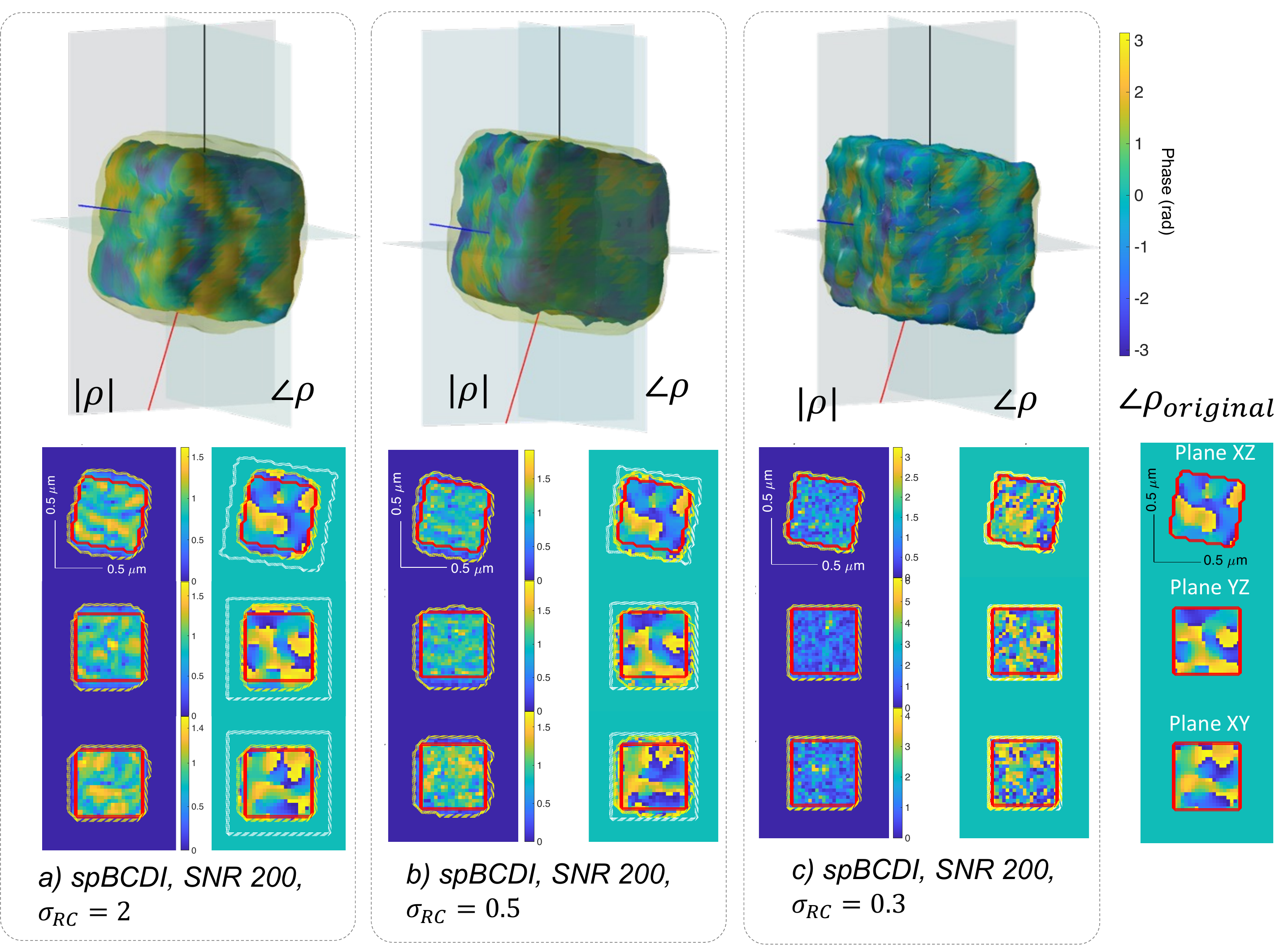}
\caption{3D reconstructions for spBCDI data sets when SW is used (\emph{i.e.}, the exact support is unknown), for $\theta_B = 9^{\circ}$. For each reconstruction, the modulus, $|\rho|$, and the phase, $\angle \rho$, of the object are shown. The initial support contour is indicated by a white dotted line, while the yellow line delineates the finally obtained support. It is also displayed in the 3D pictures with a semi-transparent yellow iso-surface. Finally, the exact support contour is marked in red.  The red, black and blue arrows indicate respectively $\kb_i$, $\kb_f$ and $\Qb_{HKL}$. Each column contains a label, which specifies the RC sampling frequency, $\sigma_{RC}$, and the SNR value.}
\label{fig:resuls_3D_9deg_OSRs_SW}
\end{center}
\end{figure*}

\paragraph{Impact of a partially known support:} The performance of the spBCDI approach when the support is unknown is also analyzed. Fig. \ref{fig:resuls_3D_9deg_OSRs_SW} shows a series of 3D reconstructions obtained when  the initial support is not perfectly known. By incorporating the shrink-wrapp (SW) routine \cite{Marchesini:2003aa}, the capability  of the reconstruction algorithm to retrieve  the phase and the shape of the cubic crystal from a spBCDI data-set is further examined. The initial support size is set to 1.5 $\times$ the object size, and its contour is shown in white in the phase cuts. Successful reconstructions are obtained for spBCDI data-sets of $\sigma_{RC} \geq 0.5$, as in the previous known support case. Note that, for $\sigma_{RC} = $ 0.5, we need to slightly reduce the initial support size to 1.3 $\times$ the object size. Thus, provided that we have a partial knowledge of the support, gathered from previous measurements at higher $\sigma_{RC}$ or other images from electronic microscopy or AFM, the ER/SW algorithms can be used successfully with strongly under-sampled spBCDI data-sets, with similar performances as the ones discussed in the previous paragraph.

\begin{figure*}
    \centering
    \includegraphics[width=1.0\linewidth]{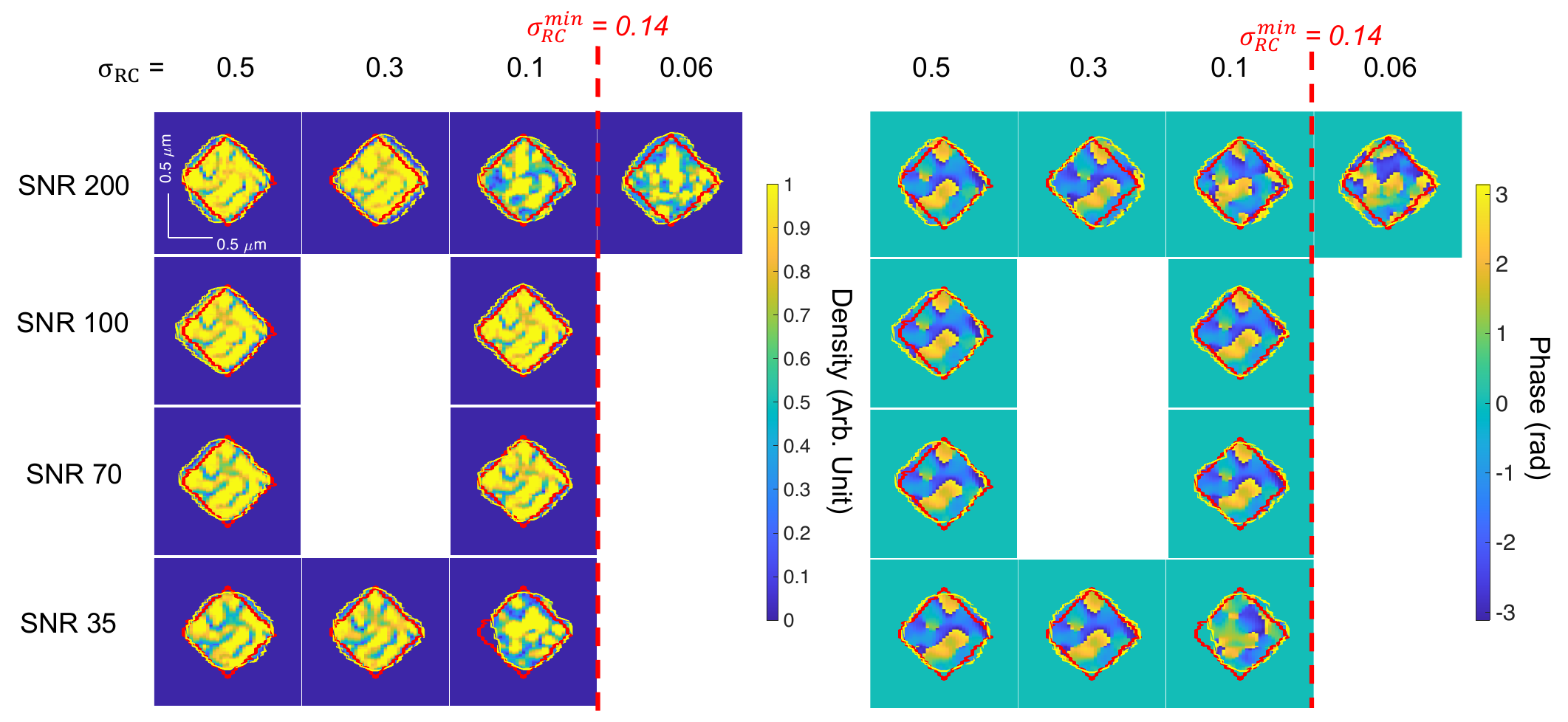}
    \caption{3D reconstructions for spBCDI data sets when SW is used (\emph{i.e.}, the exact support is unknown), for $\theta_B = 45^{\circ}$. For each reconstruction, the modulus, $|\rho|$, and the phase, $\angle \rho$, of the object are shown. The yellow line delineates the contour of the finally obtained support to be compared to the exact support contour, marked in red. The used $\sigma_{RC}$ (columns) and the SNR levels (lines) are specified in the figure.}
    \label{fig:results_45deg_lowOSR_lowSNR}
\end{figure*}

\paragraph{Impact of the incidence angle:} The other extreme case, \emph{i.e.}, the large incident angle configuration, is herein investigated. In this new series of tests (summarized in Fig. \ref{fig:results_45deg_lowOSR_lowSNR}), an incident angle of $45^{\circ}$ is chosen. In order to allow the exploration of drastically reduced $\sigma_{RC}$, the direct space pixel size along the RC direction is decreased by a factor of 3 (yielding a pixel size of $\sim$ 10 nm) while the total extension of the numerical window is preserved.  To do so, the object is interpolated in the new grid and its FT yields the 3D diffracted field whose modulus square corresponds to the 3D diffracted intensity. The later has three times more  diffraction patterns along the RC direction without affecting the underlying RC sampling rate $\delta_q$.  Then, the same definition of SNR is applied, to produce a data-set with a comparable total number of photons to the previous ones.
Thereby, even with small $\sigma_{RC}$ values, the down-sampled data-set still contains DPs to properly guide the algorithm. 
%
Roughly, for $\sigma_{RC} = 2$, the total number of diffraction patterns for one full spBCDI data-set is about 800. Results presented in Fig. \ref{fig:results_45deg_lowOSR_lowSNR} show that spBCDI still succeeds to provide reliable reconstructions for $\sigma_{RC}$ down to 0.1, corresponding to the theoretical limit given by Eq. \ref{eq:OSR_min}. In comparison with BCDI, it corresponds to a gain of about 11, comparable to the gain observed at smaller incidence angle (Table \ref{tab:DP_number}).  

\paragraph{Impact of SNR:} 

The wide angle configuration  (\emph{i.e.}, the $45^{\circ}$ incident angle geometry) is further used to provide an interesting exploration of the limits of the method with respect to very low SNR and $\sigma_{RC}$ levels. 
%
%
In this last series of tests, SNRs ranging from 100 down to 35 counts are introduced. As observed from the main results presented in Fig. \ref{fig:results_45deg_lowOSR_lowSNR}, spBCDI still performs successfully down to the theoretical limit of $\sigma_{RC} = 0.1$ for SNRs as small as 70 counts. It starts degrading at lower SNRs ($\leq$ 35 counts), which requires an increase of the $\sigma_{RC}$  value (up to 0.3) to achieve a reasonable reconstruction of the phase. In these last tests, the best achievable gain, with respect to BCDI, is of the order of 32 (Table \ref{tab:DP_number}). This means that the acquisition time can be drastically reduced, extending significantly the potential of the method for experimental applications.

\section{Discussion}
This work presents a new approach for performing BCDI microscopy with a speckle illumination. The structured beam, which produces a mix of information along the RC direction allows to overcome two important limits of traditional BCDI: (i) the need for high sampling frequencies along the RC scanning direction to ensure the robustness of the phase reconstruction for highly non-homogeneously strained crystal; (ii) the long acquisition times which precludes the imaging of time evolving  crystals. To show the performance of the new method, we have focused on a symmetric Bragg reflection, to ensure the simplicity of the simulations. But the same results are expected for a non-symmetric reflection, as long as the detector plane is significantly non-perpendicular to the beam direction (\emph{i.e.} for angles $\geq$ 10 $^\circ$). Our results confirm that the minimum sampling ratio is decreasing for larger incident angle and smaller speckle size, as established in Eq. \ref{eq:OSR_min}. Furthermore, the use of spBCDI allows working at extremely small SNR, showing that an efficient use of the diffracted photons is obtained with speckle illumination. Typically, a reduction of a factor of about 50 in SNR is observed between BCDI and spBCDI. Thereby, although spBCDI requires the acquisition of multiple RC scans performed at slightly shifted positions to avoid zero photon deposit regions, the gain with respect to acquisition time and the performances of spBCDI with respect to distorted crystals are large enough to justify its experimental implementation.\\

\indent
Regarding the gain with respect to the sampling direction, which is directly related to the total acquisition time, we observe that for individual speckle sizes of $\sim$ 50 nm, $\sigma_{RC}$ can be reduced down to 0.1 (resp 0.5) for an incident angle of 45° (resp. 9°), while BCDI requires a $\sigma_{RC}$ value of at least 2. This corresponds to a gain in sampling frequency of 20 (resp. 4) along the rocking curve. Thus, the acquisition time diminishes by a factor of 5 (resp. 1, i.e. no gain), once the need for at least four shifted acquisitions is taken into account. However, this first estimate does not account for the possibility to drastically reduce the SNR level, as observed in Fig. \ref{fig:results_45deg_lowOSR_lowSNR} and Table \ref{tab:DP_number}. Indeed, a gain of about 32 is reached for an SNR as small as 70 counts, to be compared to the needed SNR of 2000 counts for BCDI, in the case of highly distorted crystal. Transposing these numbers to experimental parameters provide an estimate of the typical accessible time scales at 4th generation synchrotrons. In a previous work \cite{Li:2022wf}, a typical acquisition time of about 33 ms for an SNR of 280 counts was reported. It would correspond to a total acquisition time of 11 s for a BCDI data set of 45 diffraction patterns and an SNR of 2000 counts, as introduced in the present work. For the low SNR regime, this acquisition time may decrease down to about 0.3 s for spBCDI. Thus, spBCDI will highly benefit from the very small (e.g. $\sim$ 30 - 50 nm) and brilliant beams generated at new 4th generation light sources \cite{Raimondi:2023ub,Tavares:xe5034}. Note, that this preliminary estimation does not account for the time needed to move the different scanning motors (specially the rocking motors), which will become limiting factors for temporal resolution. However, we anticipate that additional gains could be obtained by shifting the angular positions by less than one angular step or to use  continuous rotation schemes based on flyscans \cite{Pelz:2014aa}, for each of the four RC. This partial overlapping in reciprocal space should still contain enough information to allow for additional reduction of the sampling conditions with speckle beams.\\

\indent
Regarding the gain for imaging complex crystalline samples, the present test would correspond to a strain distribution with a total extent of almost 10$^{-2}$, with most of the non-homogeneous strains being included in a range of about 10$^{-3}$ and a small portion of the strains reaching an amplitude of a few 10$^{-3}$ (see Supplementary Information). Such a strain distribution compares well to experimental strain fields in experimental works by Ulvestad et al. \cite{Ulvestad:2015aa} in the study of topological defects in battery grains  or by Sun et al. \cite{Sun_Singer:2021to} within single grains in solid-state electrolytes.These works reported strain amplitudes of $4\cdot10^{-3}$ and even larger ($ 7\cdot10^{-3}$). Remarkably, spBCDI keeps on performing well even at low SNR, which is not the case for BCDI, as observed when reducing the SNR by a factor of 2, from 2000 counts to 1000 counts (Table \ref{tab:DP_number}). \\

\indent
While the scope of this work is to numerically establish the relevance of spBCDI, the experimental demonstration is desirable. Practically, the implementation of the proposed approach requires to discuss the needs for the phase plate production, the coupling of the phase plate with the focusing optics, the scanning stage and the diffractometer. The phase plate can be produced by two-photon lithography \cite{Seiboth:2020aa,Seiboth:2022aa} of an x-ray phase shifting polymer. The typical size of domains which present rather homogeneous phase directly drives the beam envelope size. Our estimation, presented in Fig. \ref{fig:MPP_spbeam}, shows that these domains, in the 1-10 micrometer range, are compatible with present printing technologies. When needed, the domain size can be easily adapted to the crystal size, ensuring that most of the photons at the focal plane illuminate the sample. This strategy is in contrast to standard BCDI, based on  plane-wave illumination and where only the central part of the beam is used.\\
The phase plate, which shapes the illumination, should be installed close to the focusing optics. Thereby, it benefits from the stable environment surrounding  the optics. As a result, the beam produced by the combination of the phase plate and the focusing optics is weakly sensitive to their relative vibration, as the wave-field close to the focusing element is only weakly structured. The phase plate positioning, alignment and characterization can be performed on the  forward geometry, without the need for the crystalline sample.\\
Besides the specificity regarding the illumination, the spBCDI set-up is rather similar to the Bragg ptychography set-up, which is available at several synchrotrons, worldwide \cite{Hruszkewycz:2017ua,Li:2021ud,Li:2022wf,Mastropietro:2017un,Berenguer_Chamard:2013}. This includes a scanning translation stage and a diffractometer for crystalline alignment without further modification.

Finally, we note that Zhao et al. \cite{Zhao:2023aa} have recently proposed to perform BCDI with a modulator, presenting some similarities with our spBCDI method. Their approach, which aims at pushing further the capabilities of BCDI for hihgly distorted crystalline samples, consists in placing the modulator after the sample and parallel to the detector plane. Thereby, they increase substantially the spatial resolution achieved in each slice of the RC scan, yielding data-sets with sufficient constraints to produce a robust and unique image of the crystal phase domains. However, because the modulation of the exit-field does not mix the information along the RC scan, the ability to relax the sampling conditions is not expected. This is a key difference with our proposed spBCDI. The practical implementation is also rather different as the phase plate should be placed close to the sample, involving additional constraints with respect to spBCDI.

\section{Summary and conclusions}
This paper proposes a new methodology in the framework of coherent x-ray microscopy, spBCDI, based on a strongly non-uniform illumination or \emph{speckle illumination}. Thereby, we increase the efficiency with which the reciprocal space is sampled and drastically diminish the measurement times. We describe the theoretical foundations of the new method, predicated on the definition of the diffracted wave field as a convolution between the object and the illumination FT. By altering the illumination, we produce shifted copies of the physically observable region (\emph{i.e.} the Bragg peak) and enable the sampling of extra information, which would be inaccessible with plane wave illumination. Thus, speckle illumination produces a multiplexing of the Bragg peak, and hence, a noticeable reduction in the oversampling ratio along the RC scan, $\sigma_{RC}$. Besides, we provide with a qualitative rule which connects the experimental parameters (\emph{i.e.} incident angle and speckle size) to the minimum $\sigma_{RC}$ above which the data-set can be inverted, yielding a robust and unique reconstruction of the object's phase. Numerical investigations show that values of $\sigma_{RC}$ as small as 0.1 could be reached, to be compared to the Nyquist-Shannon requirement of $\sigma_{RC} =2$ for BCDI. The observed reduction in SNR leads to a decrease of total acquisition time by a factor of 30. Finally, some details regarding the practical implementation of the method at synchrotron sources are discussed, including the experimental methodology to design a speckle illumination, fully characterized by a specific speckle size and an envelop determining the field of view.\\

\indent
We conclude that spBCDI offers an attractive alternative approach to traditional plane wave BCDI, which combines faster measurement with the ability of imaging strongly distorted crystals, even at low SNR levels. Typically at 4th generation synchrotron sources, total acquisition times in the order of 0.3 s could be reached. Our next step is to apply spBCDI to dynamic crystalline systems. We believe that the possibility of imaging dynamic crystals will appeal a wide variety of scientific communities, including physicist, chemist, geologists, biologists or materials scientists. Finally, the new capabilities of 4th generation light sources will promote these techniques to observe the evolution of crystalline matter \emph{in situ} and under realistic conditions. Therefore, we expect that spBCDI will be readily adopted in these new radiation sources, which are becoming more and more available all over the world.

\section{Acknowledgements} 
I. C-A acknowledges funding by the European-Next-GenerationEU action. She also acknowledges  financial support from the Spanish Agencia Estatal de Investigaci\'on, through projects PID2020-115159GB-I00/AEI/10.13039/501100011033, Aragonese project RASMIA E12-23R co-funded by Fondo Social Europeo of the European Union FEDER (ES). The authors also want to thank A. Kubec from XRnanotech GmbH, Switzerland, for extensive discussions regarding the feasibility of the phase plate modulator production, based on specific designs fulfilling the needs of spBCDI .

\pagebreak

\bibliography{./biblio_spBCDI_sim} 
\bibliographystyle{apsrev4-2} 

\end{document}